\documentclass[12pt,preprint]{aastex}

\shorttitle{Unresolved Canceling Mangetic Elements}
\shortauthors{Kubo, Low, \& Lites}

\begin{document}


\title{Unresolved Mixed Polarity Magnetic Fields at Flux Cancellation
Site in Solar Photosphere at $0\arcsec.3$ Spatial Resolution}  


\author{Masahito Kubo}
\affil{National Astronomical Observatory, Mitaka, Tokyo, 181-8588, Japan.} 
\email{masahito.kubo@nao.ac.jp}

\author{Boon Chye Low and Bruce W Lites}
\affil{High Altitude Observatory, National Center for
Atmospheric Research\altaffilmark{1}, P.O. Box 3000, Boulder, CO 80307, USA.}


\altaffiltext{1}{The National Center for Atmospheric Research is sponsored by the National Science Foundation}

\begin{abstract}
This is a follow-up investigation of a magnetic-flux cancellation event
 at a polarity inversion line (PIL) on the Sun observed with the
 spectropolarimeter on board \textit{Hinode}.
Anomalous circular polarization (Stokes \textit{V}) profiles are
 observed in the photosphere along the PIL at the cancellation sites. 
\citet{Kubo2010} previously reported that the theoretically expected
 horizontal fields between the canceling opposite-polarity magnetic
 elements in this event are not detected at granular scales. 
We show that the observed anomalous Stokes \textit{V} profiles are
 reproduced successfully by adding the nearly symmetric Stokes
 \textit{V} profiles observed at pixels immediately adjacent to the 
 PIL. 
This result suggests that these observed anomalous Stokes \textit{V}
 profiles are not indications of a flux removal process, but are the
 result of either a mixture of unresolved, opposite-polarity magnetic
 elements or the unresolved width of the PIL, at an estimated resolution 
 element of about 0''.3.  
The hitherto undetected flux removal process accounting for the
 larger-scale disappearance of magnetic flux during the observing period
 is likely to also fall below resolution.
\end{abstract}


\keywords{Sun: magnetic fields --- Sun: photosphere --- line: profiles --- techniques: polarimetric}

\section{Introduction}
We address the mutual disappearance of opposite-polarity magnetic
 elements from the solar photosphere following their apparent
 ``collisions,'' as observed in line-of-sight magnetograms. 
This phenomenon is called ``magnetic flux cancellation'' \citep{Martin1985}.
The study of magnetic flux cancellation is important for understanding
 the nature of the flux removal process from the solar surface layers. 
\citet{Kubo2010} investigated five flux cancellation events at granular
scales in which the horizontal magnetic fields between the canceling
 opposite-polarity magnetic elements were detected in only one 
event that takes place in a small emerging flux region.
This finding is interesting since almost all theoretical
 scenarios proposed to explain photospheric flux cancellation expect an
 increase in the horizontal magnetic field between the canceling
 opposite-polarity magnetic elements \citep[e.g.][]{Zwaan1987}.
Magnetic reconnection in the photosphere or around the temperature
 minimum region has been proposed to explain the formation of the horizontal
 magnetic field \citep{Litvinenko1999, Takeuchi2001, Ryutova2003}.
Such horizontal magnetic fields have been observed in some events
 \citep{Chae2004, Kubo2007, Iida2010}.
Prior to the work of \citet{Kubo2010}, only one flux cancellation event
 without a detection of horizontal magnetic field was reported
 \citep{Bellot2005}.
The increase in horizontal magnetic field expected theoretically has not
 been commonly observed at least in flux cancellation events at granular
 scales \citep{Kubo2010}. 
The Solar Optical Telescope \citep[SOT;][]{Tsuneta2008} on board
\textit{Hinode} \citep{Kosugi2007} has shown that these
 small-scale flux cancellation events at granular scales may be observed
 everywhere on the solar surface which clearly makes this
 phenomenon key to understanding the removal of photospheric magnetic flux. 
\citet{Kubo2010} suggested that flux cancellations with and without
 detectable horizontal fields are physically distinct processes.
They also cautioned that the SOT observations of flux removal events at
 granular scales without the appearance of the horizontal fields may
 not be spatially resolved.
This caution is based only on the null detection of linear polarization
 signals representing horizontal magnetic fields.
However, circular polarization signals (Stokes \textit{V}) are also
 detected at the polarity inversion line in these events.
In the present paper we investigate the observational implications of
 the observed Stokes \textit{V} profiles for the spatially
 unresolved flux removal process. 

The Doppler velocity at the flux cancellation site is an important
parameter that may discriminate between an 
emerging U-loop model or a submerging ${\Omega}$-loop model \citep[see
Figure~2 in][]{Zwaan1987}. 
The zero-crossing wavelength of the Stokes \textit{V} profile is
sometimes used to derive a Doppler velocity of the magnetized
atmosphere. 
However, the Stokes \textit{V} profiles observed along the
polarity inversion line often take asymmetric or multi-lobed forms 
\citep[e.g,][]{Sigwarth2001}.
It is difficult to determine the zero-crossing wavelength of
such anomalous Stokes \textit{V} profiles. 
There are two processes commonly invoked to explain anomalous Stokes
\textit{V} profile: (1) a coupling of gradients in velocity and 
magnetic field along the line-of-sight (LOS)
\citep[e.g.,][]{Illing1975, Auer1978, Sanchez1992, Solanki1993},
or (2) the superposition of two or more magnetic components of
both polarities with relative Doppler shifts within a resolution element 
\citep[e.g.,][]{Grigorjev1972, Golovko1974, Sanchez1992}.
A correct interpretation of the anomalous Stokes \textit{V} profiles
is essential for interpreting the magnetic field structures in
the neighborhood of the polarity inversion line between the canceling
magnetic elements.

\section{Observations}
We investigate circular polarization (Stokes \textit{V}) profiles at the
polarity inversion line formed by canceling opposite-polarity magnetic
elements just outside an active region NOAA 10944 on 2007 March 2.  
The SOT spectropolarimeter \citep[SP,][]{Lites2013a} provides the
Stokes profiles with a $0\arcsec.15$ slit width and a $0\arcsec.16$
pixel sampling along the slit.
The Stokes profiles were measured with an integration time of
4.8 s at each slit position. 
The calibration of the Stokes profiles was done with a standard
procedure \citep{Lites2013b}.

\begin{figure}
\epsscale{1.0}
\plotone{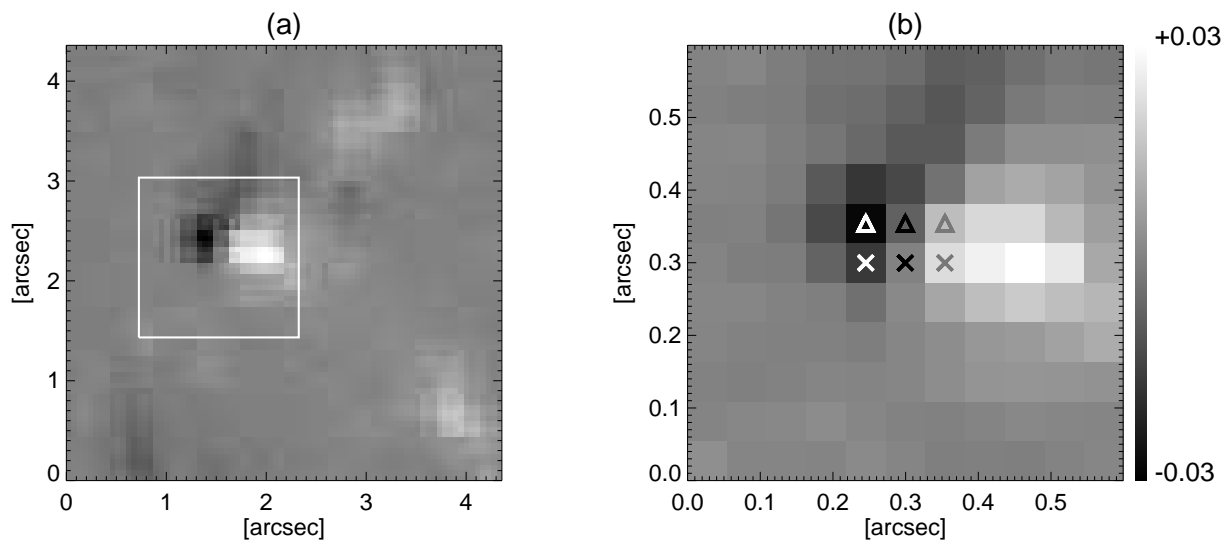}
\caption{
Total circular polarization ($C_{tot}$) map for the whole flux
 cancellation region (panel \textit{a}), and for the canceling magnetic
 bipole (panel \textit{b}).  
The map was constructed from a spectrograph scan executed between
 09:38:36 to 09:44:04 on 2007 March 2.
Panel \textit{a} is same as the third frame of Figure~3a in
 \citet{Kubo2010}.
The solid box of panel \textit{a} is identical to the field of view
 of panel \textit{b}. 
The black symbols in panel \textit{b} represent pixels at the polarity
 inversion line. 
The white and gray symbols represent pixels next to the polarity
 inversion line, and they have negative and positive magnetic polarity,
 respectively.
}
\label{fig_sfov1}
\end{figure}

The total circular polarization ($C_{tot}$) is estimated as follows:
\begin{equation}
C_{tot}= \frac{\int_{\lambda_0-21.6\thinspace\textrm{\scriptsize{pm}}}^{\lambda_0-4.32\thinspace\textrm{\scriptsize{pm}}}V(\lambda) \,d\lambda}{I_c\int_{\lambda_0-21.6\thinspace\textrm{\scriptsize{pm}}}^{\lambda_0-4.32\thinspace\textrm{\scriptsize{pm}}}\,d\lambda}.
\end{equation}
The center of the Fe \small{I} 630.25 nm line ($\lambda_0$) is defined in
each pixel as the center of gravity of the Stokes \textit{I} profile. 
The local continuum intensity ($I_c$) is defined as the average of the
Stokes \textit{I} profile from $\lambda_0+43.2$ pm to $\lambda_0+64.8$ pm.
Figure~\ref{fig_sfov1} shows a map of the total circular polarization
for our target cancellation event.
The opposite-polarity magnetic elements independently appear in the
quiet area outside the moat region and then approach each other.
The cancellation activity of these magnetic elements is observed during
a 10 minute period after the map of Figure~\ref{fig_sfov1}, but
the linear  polarization signal does not increase near the polarity
inversion line during the cancellation. 
The temporal evolution and detailed properties of this cancellation event 
were described as ``region A'' in \citet{Kubo2010}. 

Doppler shifts are derived both from $\lambda_0$ of the Stokes
\textit{I} profile and from the zero-crossing wavelength of the Stokes
\textit{V} profile. 
A reference wavelength to the Doppler shift is the average of
$\lambda_0$ of Stokes \textit{I} profiles in the quiet area, and the same
value is used as the reference wavelength for the zero-crossing
wavelength of Stokes \textit{V} profiles. 
The zero-crossing wavelength is derived by the linear fit to successive
four wavelength points crossing the zero.

\section{Results}
\begin{figure}
\epsscale{1.0}
\plotone{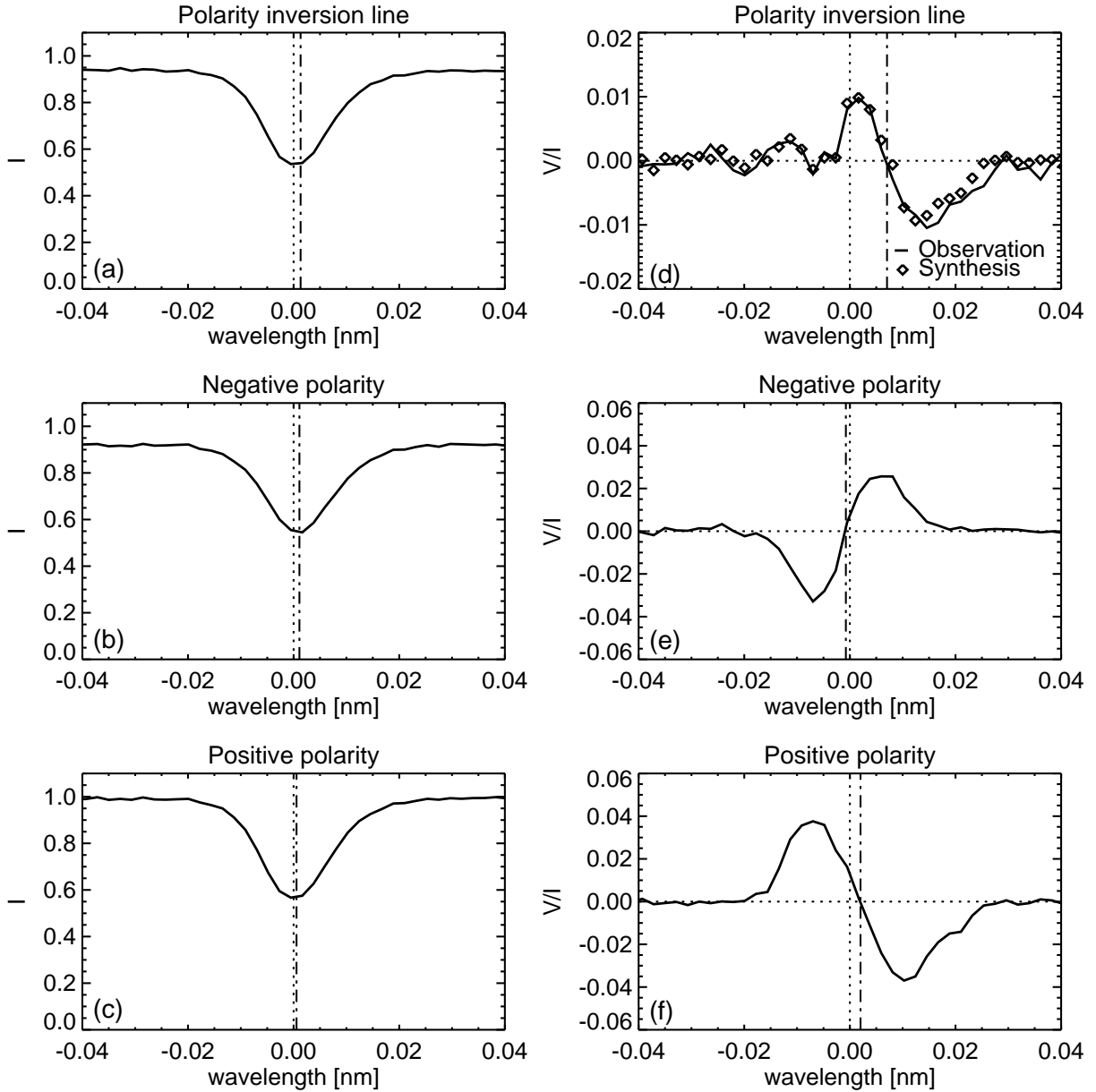}
\caption{
Panels \textit{a}, \textit{b}, and \textit{c} show the observed Stokes
 \textit{I} profiles of the Fe \small{I} 630.25 nm line 
 at pixels represented by black, white, and gray cross symbols in panel
 \textit{b} of Figure~\ref{fig_sfov1}, respectively.
These profiles are normalized by the continuum intensity averaged over
 the quiet area. 
The solid lines in panels \textit{d-f} are observed Stokes \textit{V}
 normalized by the continuum intensity at the pixels same as their
 left panels.  
The diamond symbols in panel \textit{d} show the profile synthesized
 from the observed profiles in panels \textit{e} and \textit{f}.  
The vertical dotted line represents the averaged position of the line
 centers over the map. 
The vertical dash-dotted line represents a line center for the Stokes \textit{I}
 profile and a zero-crossing wavelength for the Stokes \textit{V}
 profile.
} 
\label{fig_profile1}
\end{figure}

\begin{figure}
\epsscale{1.0}
\plotone{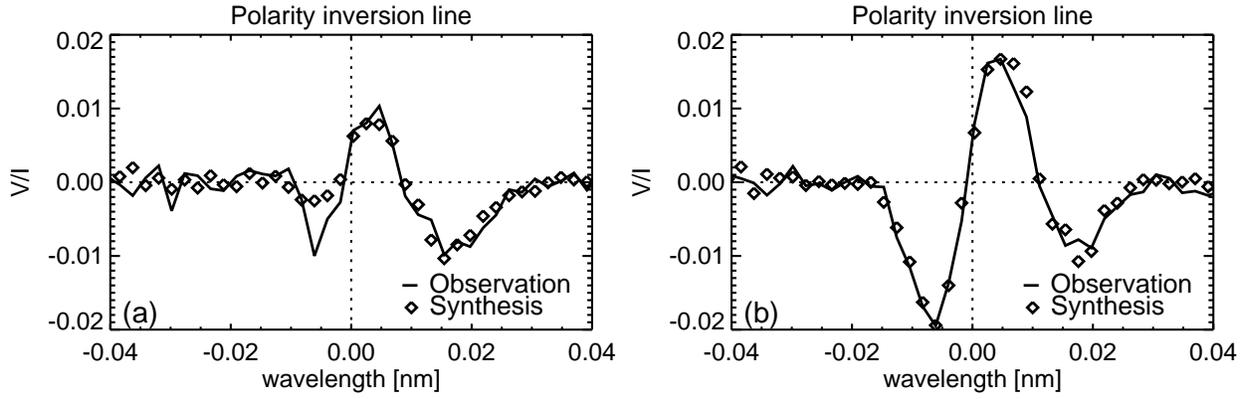}
\caption{
Panel \textit{a} is same as panel \textit{d} of Figure~\ref{fig_profile1}
 but for Stokes \textit{V} profile of the Fe \small{I} 630.15 nm line.
Panel \textit{b} is same as panel \textit{a} but for the pixel
 represented by the black triangle symbol in panel \textit{b} of
 Figure~\ref{fig_sfov1}.
}
\label{fig_profile6301}
\end{figure}

A highly asymmetric, strongly redshifted Stokes \textit{V} profile
in the Fe \small{I} 630.25 nm line is observed at the polarity
inversion line, as shown in Figure~\ref{fig_profile1}\textit{d}. 
If we interpret the shift of the zero-crossing wavelength as
being caused by a Doppler velocity, it would be 3.3 km s$^{-1}$, which
is much higher than the Doppler velocity of the Stokes \textit{I}
profile (-0.6 km s$^{-1}$).
The three-lobed shape is more clearly seen in the Stokes
\textit{V} profile of Fe \small{I} 630.15 nm
(Figure~\ref{fig_profile6301}\textit{a}), and one of two zero-crossing
wavelength positions is more redshifted for the Fe \small{I} 630.25nm
line.  
Hereafter, we try to explain such a strange Stokes \textit{V} profile by
mixing profiles of the canceling magnetic elements.
The Stokes \textit{V} profile at the polarity inversion line
($V_{PIL}^{syn}$) is synthesized as follows: 
\begin{equation}
\label{Synthesis_V}
V_{PIL}^{syn} = A\times V_{+}^{obs} + B\times V_{-}^{obs},
\end{equation}
\noindent
where $V_{+}^{obs}$ and $V_{-}^{obs}$ are the Stokes
\textit{V} profiles observed at the pixels next to the polarity
inversion line on the positive polarity side and on the negative
polarity side, respectively (Figures~\ref{fig_profile1}\textit{e} and
\ref{fig_profile1}\textit{f}). 
Note that nearly symmetric Stokes \textit{V} profiles are observed
at the pixels next to the polarity inversion line.
The positive constant values of $A$ and $B$ are determined to minimize
the difference between the observed Stokes \textit{V} profile and 
the synthesized profile at the polarity inversion line inside a
wavelength range of $\pm$ 43.1 pm from the averaged center of the Fe
\small{I} 630.25 nm line.
The values calculated from this fit are $A = 0.42$ and $B = 0.52$.
The synthesized Stokes \textit{V} profile is shown by the diamond
symbols in Figure~\ref{fig_profile1}\textit{d}, and it looks very
similar to the observed one.  
Most of local dips and peaks in the Stokes \textit{V} profile at the
polarity inversion line are successfully demonstrated in spite of such a
simple summation of the observed Stokes \textit{V} profiles.  
\citet{Beck2008} performed a similar addition of opposite-polarity
Stokes \textit{V}  profiles to reproduce multi-lobed Stokes
\textit{V} profiles like those within sunspot penumbrae. 
He stressed the importance of relative Doppler shifts between the
opposite polarity elements. 
In the present work, we infer a relative Doppler velocity of 1.3 km
s$^{-1}$ between the pixels next to the polarity inversion
line, but we find a relative difference of -0.3 km s$^{-1}$ for the
corresponding Stokes \textit{I} profiles.
The different spectral extents of the Stokes \textit{V} lobes in
the positive and negative-polarity magnetic elements also help in the
generation of anomalous Stokes \textit{V} profile at the polarity
inversion line.
The different spectral extents originate from the difference in
the field strength of the canceling opposite polarity magnetic
patches. 
Figure~\ref{fig_profile6301}\textit{a} shows that the
reproduction of the Stokes \textit{V} profile in the Fe \small{I} 630.15
nm line is as good as that of the Fe \small{I} 630.25 nm 
line with similar values for $A$ and $B$ ($A = 0.39$
and $B = 0.52$).

\begin{figure}
\epsscale{1.0}
\plotone{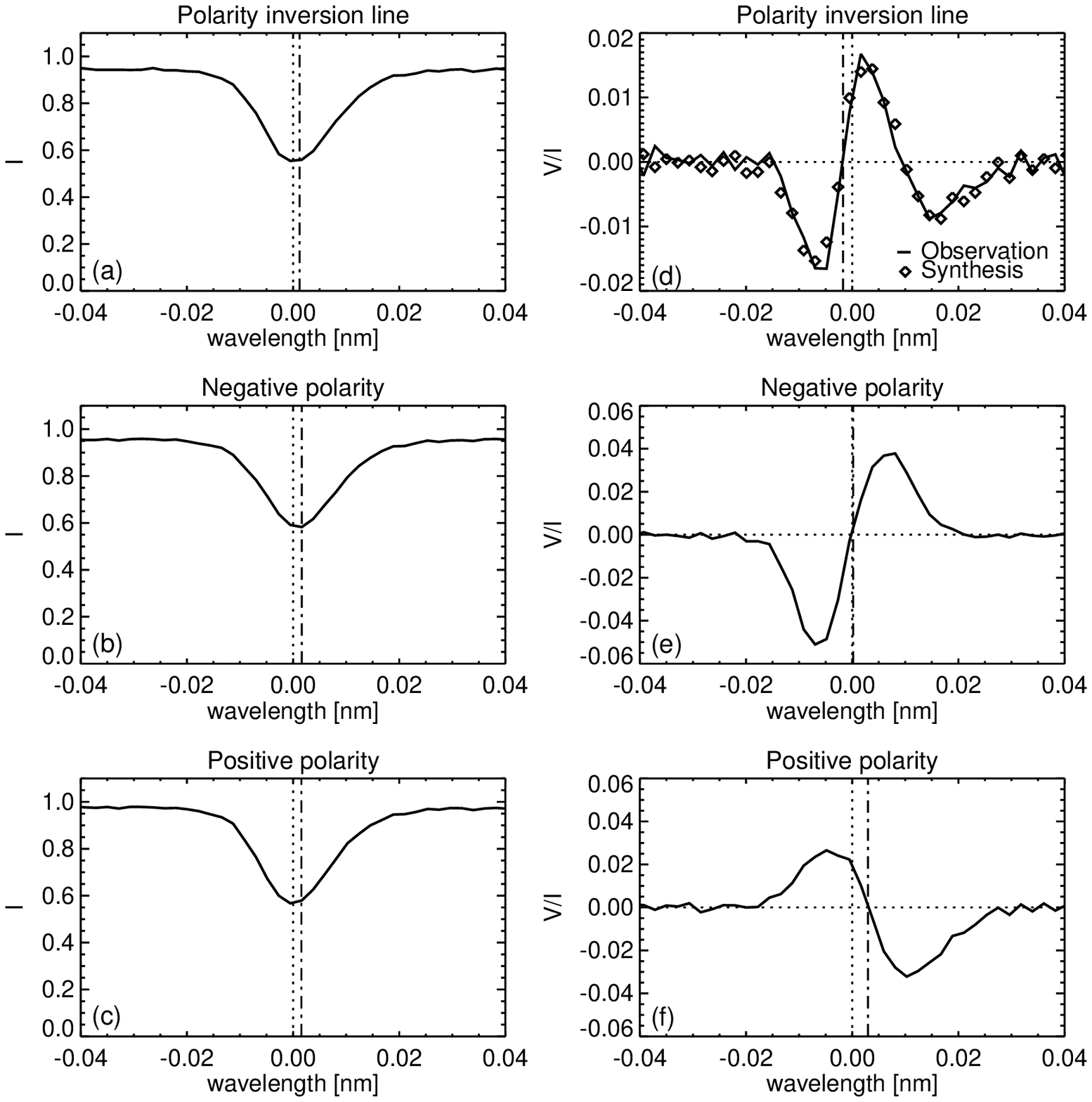}
\caption{
Same as Figure~\ref{fig_profile1}, but for the pixels represented by the
 triangle symbols in panel \textit{b} of Figure~\ref{fig_sfov1}.
}
\label{fig_profile2}
\end{figure}

Similar results are obtained at the neighboring pixel along the
polarity inversion line, as shown in Figures~\ref{fig_profile2}\textit{d-f}.
A three-lobed profile is observed at the polarity inversion
line, which is produced by the sum of the observed profiles at
the pixels next to the polarity inversion line.
In this case, the coefficients $A$ and $B$ are 0.52 and 0.54 for the
Stokes \textit{V} profile in the Fe \small{I} 630.25nm line, respectively.
These values are similar to those of the previous case.
The value of $A$ is 0.49 and $B$ is 0.61 for the Fe \small{I}
630.15nm line. 
The different coefficients between the two Fe \small{I} lines may be
caused by the height gradient of the magnetic field structures at the
pixel of polarity inversion line or at the neighboring pixels.

\section{Discussion}
Highly asymmetric or three-lobed Stokes \textit{V} profiles are
observed at the polarity inversion line between the canceling magnetic
elements.
Such anomalous Stokes \textit{V} profiles at the polarity inversion
line are successfully produced by the sum of the Stokes \textit{V} profiles
observed at the pixels next to the polarity inversion line. 
This could serve to warn of over-interpreting Stokes profiles
observed at the polarity inversion line.
For example, the Doppler velocity derived from the zero-crossing
wavelength of the Stokes V profile is not necessarily reliable.
The mixture of opposite-polarity Stokes \textit{V} profiles with a
relative Doppler shift causes the anomalous Stokes \textit{V} profiles
at the polarity inversion line formed by canceling opposite-polarity
magnetic elements like those observed in the penumbra \citep{Beck2008}.
The large relative Doppler shift is not observed in the Stokes
\textit{I} profiles, but is observed in the Stokes \textit{V} profiles at
the pixels next to the polarity inversion line.

The combination of opposite-polarity Stokes \textit{V} profiles
could originate from the presence of unresolved
opposite-polarity magnetic elements or the unresolved boundary between
the opposite polarity patches. 
We cannot investigate the mixture of magnetic fields within the pixel
because we analyze Stokes profiles in a single pixel of 0$\arcsec$.16
width which is about half of a diffraction limit of a 50 cm diameter
telescope at 630.25 nm.
The diffraction-limited performance of the telescope with a
Strehl ratio of about 0.8 was confirmed with images taken by the
Broadband Filter Imager of SOT \citep{Suematsu2008, Wedemeyer-Bohm2008}.
The rms contrast of the continuum intensity is 8.1$\%$ in our
data set, which is similar to the rms contrast (8.5$\%$) of a
synthetic image from the MHD simulation that was degraded by a
point-spread function of the SOT/SP without the defocus in
\citet{Danilovic2008}. 
This means that our observations are performed under the almost best
focus condition and the size of a resolution element is about 0''.32.
Although it is difficult to clearly distinguish the mixture of
unresolved opposite polarity elements from the unresolved boundary
between the opposite polarity magnetic patches,
our result may support the later case because we can reproduce
the Stokes \textit{V} profiles at the polarity inversion line using a
fraction of the neighboring profiles, i.e. there are no indications of
other magnetic elements in the area.

Magnetic flux disappears in this flux cancellation event during our
observing period \citep{Kubo2010}.  
Therefore, it is expected that a magnetic flux removal process is
actually operative in this event.
However, our results suggest that the Stokes \textit{V} profile arising
from the flux removal process is not yet detected, at least in the event
studied herein. 
The mixture of the opposite polarity magnetic elements to be canceled is 
limited within a local area around the polarity inversion line
since nearly symmetric profiles are observed at the pixels next to the
polarity inversion line. 
One possibility of a lack of the Stokes \textit{V} profile arising from the
flux removal process is that the spatial resolution of the SOT/SP
($\sim$200 km) is still insufficient to detect the removal process of
photospheric magnetic flux \citep{Kubo2010}.
Whether it is possible to observe the horizontal fields expected in the
flux-removal process of the event studied depends on both observational
resolution as well as the actual MHD nature of this process.  
The latter is worthy of theoretical investigation but lies outside the
scope of the paper.
For the present there is much to learn from Stokes-polarimetric
analysis of photospheric events involving the interaction of flux
elements of opposite magnetic polarities such as presented in our paper
and elsewhere \citep[e.g.][]{Rezaei2007}.

\acknowledgments
We gratefully acknowledge the helpful comments and suggestions
of an anonymous referee.
We also thank Y. Suematsu for discussions on the point-spread 
function of the \textit{Hinode} SOT, and Y. Katsukawa and
K. Ichimoto for useful discussions on this paper.
\textit{Hinode} is a Japanese mission developed and launched by 
ISAS/JAXA, with NAOJ as domestic partner and NASA and STFC (UK) as
international partners. It is operated by these agencies in cooperation
with ESA and NSC (Norway). 
The FPP project at LMSAL and HAO is supported by NASA contract NNM07AA01C.


\begin{thebibliography}{}

\bibitem[Auer \& Heasley(1978)]{Auer1978} Auer, L.~H., \& Heasley, J.~N.\ 1978, \aap, 64, 67 

\bibitem[Beck(2008)]{Beck2008} Beck, C.\ 2008, \aap, 480, 825 
\bibitem[Bellot Rubio \& Beck(2005)]{Bellot2005} Bellot Rubio, L.~R., \& Beck, C.\ 2005, \apjl, 626, L125 

\bibitem[Chae et al.(2004)]{Chae2004} Chae, J., Moon, Y., \& Pevtsov, A.~A.\ 2004, \apjl, 602, L65 

\bibitem[Danilovic et al.(2008)]{Danilovic2008} Danilovic, S., Gandorfer, A., Lagg, A., et al.\ 2008, \aap, 484, L17 

\bibitem[Golovko(1974)]{Golovko1974} Golovko, A.~A.\ 1974, \solphys, 37, 113 
\bibitem[Grigorjev \& Katz(1972)]{Grigorjev1972} Grigorjev, V.~M., \& Katz, J.~M.\ 1972, \solphys, 22, 119 

\bibitem[Iida et al.(2010)]{Iida2010} Iida, Y., Yokoyama, T., \& Ichimoto, K.\ 2010, \apj, 713, 325 
\bibitem[Illing et al.(1975)]{Illing1975} Illing, R.~M.~E., Landman, D.~A., \& Mickey, D.~L.\ 1975, \aap, 41, 183 


\bibitem[Kosugi et al.(2007)]{Kosugi2007} Kosugi, T., et al.\ 2007, \solphys, 243, 3 
\bibitem[Kubo et al.(2010)]{Kubo2010} Kubo, M., Low, B.~C., \& Lites, B.~W.\ 2010, \apj, 712, 1321 
\bibitem[Kubo \& Shimizu(2007)]{Kubo2007} Kubo, M., \& Shimizu, T.\ 2007,
		\apj, 671, 990 


\bibitem[Lites et al.(2013)]{Lites2013a} Lites, B.~W., Akin, D.~L., Card, G., et al.\ 2013, \solphys, 283, 579 
\bibitem[Lites \& Ichimoto(2013)]{Lites2013b} Lites, B.~W., \& Ichimoto, K.\ 2013, \solphys, 283, 601 
\bibitem[Litvinenko(1999)]{Litvinenko1999} Litvinenko, Y.~E.\ 1999, \apj, 515, 435 

\bibitem[Martin, Livi, \& Wang(1985)]{Martin1985} Martin, S.~F., Livi, S.~H.~B., \& Wang, J.\ 1985, Australian Journal of Physics, 38, 929 

\bibitem[Rezaei et al.(2007)]{Rezaei2007} Rezaei, R., Schlichenmaier, R., Schmidt, W., \& Steiner, O.\ 2007, \aap, 469, L9 
\bibitem[Ryutova et al.(2003)]{Ryutova2003} Ryutova, M., Tarbell, T.~D., \& Shine, R.\ 2003, \solphys, 213, 231 

\bibitem[Sanchez Almeida \& Lites(1992)]{Sanchez1992} Sanchez Almeida, J., \& Lites, B.~W.\ 1992, \apj, 398, 359 
\bibitem[Sigwarth(2001)]{Sigwarth2001} Sigwarth, M.\ 2001, \apj, 563, 1031 
\bibitem[Solanki \& Montavon(1993)]{Solanki1993} Solanki, S.~K., \& Montavon, C.~A.~P.\ 1993, \aap, 275, 283 
\bibitem[Suematsu et al.(2008)]{Suematsu2008} Suematsu, Y., Tsuneta, S., Ichimoto, K., et al.\ 2008, \solphys, 249, 197  

\bibitem[Takeuchi \& Shibata(2001)]{Takeuchi2001} Takeuchi, A., \& Shibata, K.\ 2001, Earth, Planets, and Space, 53, 605 
\bibitem[Tsuneta et al.(2008)]{Tsuneta2008} Tsuneta, S., et al.\ 2008, \solphys, 249, 167 

\bibitem[Wedemeyer-B{\"o}hm(2008)]{Wedemeyer-Bohm2008} Wedemeyer-B{\"o}hm, S.\ 2008, \aap, 487, 399 

\bibitem[Zwaan(1987)]{Zwaan1987} Zwaan, C.\ 1987, \araa, 25, 83 

\end{thebibliography}
\end{document}